\documentclass[preprint,aps,showpacs]{revtex4}
\usepackage{amsmath}

\begin{document}

\title{A pair of oscillators interacting with a common heat bath}
\author{G. W. Ford}
\affiliation{Department of Physics, University of Michigan, Ann Arbor, MI 48109-1040 USA}
\author{R. F. O'Connell\thanks{
oconnell@phys.lsu.edu}}
\affiliation{Department of Physics and Astronomy, Louisiana State University, Baton
Rouge, LA 70803-4001 USA}
\date{\today }

\begin{abstract}
Here the problem considered is that of a pair of oscillators coupled to a
common heat bath. Many, if not most, discussions of a single operator
coupled to a bath have used the independent oscillator model of the bath.
However, that model has no notion of separation, so the question of
phenomena when the oscillators are near one another compared with when they are widely
separated cannot be addressed. Here the Lamb model of an oscillator attached
to a stretched string is generalized to illustrate some of these questions.
The coupled Langevin equations for a pair of oscillators attached to the
string at different points are derived and their limits for large and small
separations obtained. Finally, as an
illustration of a different phenomenon, the fluctuation force between a pair
of masses attached to the string is calculated, with closed form expressions
for the force at small and large separations..
\end{abstract}

\maketitle

Many, if not most, discussions of a quantum oscillator interacting with a
quantum heat bath have been based on the independent oscillator mode of the
bath. A major advantage of this model is that it can be shown to be
equivalent with the most general linear passive bath, even one coupled to
multiple degrees of freedom. \cite{ford88, li90} \ However, an attempt to
construct such a model to describe two separated oscillators does not lead
to results of interest, essentially because the independent oscillator model
does not lend itself to the notion of physical separation. However, the Lamb
model of an oscillator attached to a stretched string does have the notion
of position: a pair of oscillators can be attached to separated points on
the string. This model was introduced by Horace Lamb in 1900 \cite{lamb00},
in order to better understand radiation reaction in electrodynamics. Later
it was shown by Lewis and Thomas \cite{lewis75} that the Lamb model is a
special case of the independent oscillator model. Thus we have an
independent oscillator model (the oscillators are the normal mode
oscillators of the string) in which the notion of separation is obvious. Our
aim in this paper will be to describe some of the interesting physics that
arises when a pair of oscillators are attached to separated points on the
string.

We begin by defining the model and using it to obtain a pair of coupled
quantum Langevin equations describing the motion of the pair of oscillators.
The procedure here is a straightforward generalization of that used to
obtain the equation for a single oscillator attached to the string.\cite
{ford88} \ The results, however, show some striking phenomena: interaction
of the oscillators produced by waves in the string propagating from one to
the other, correlation of the random forces acting on each oscillator. We
then show how simple descriptions of the motion arise in the limits of large
and small separations. 

As an example of a different phenomenon, we then turn to a calculation of
the fluctuation force acting between a pair of masses attached to the
string. This force is a one dimensional analog of the London-van der Waals
force between molecules or the Casimir force between parallel plates. As in
those cases, we find simple expressions for the force at small and large
separations.

The Lamb model for a pair of oscillators coupled to points on a stretched
string is described by the Lagrangian:
\begin{equation}
L={\frac{1}{2}}m_{1}\dot{x}_{1}^{2}-\frac{1}{2}K_{1}x_{1}^{2}+{\frac{1}{2}}
m_{2}\dot{x}_{2}^{2}-\frac{1}{2}K_{2}x_{2}^{2}+\int_{-\infty }^{\infty }dy[{
\frac{\sigma }{2}}({\frac{\partial u}{\partial t}})^{2}-{\frac{\tau }{2}}({
\frac{\partial u}{\partial y}})^{2}],  \label{ent1}
\end{equation}
where $u(y)$ is the string displacement. The mass per unit length of the
string is $\sigma $ and the tension is $\tau $. Note that the string is
stretched along the $y$-axis and the oscillator displacements are along the $
x$ -axis, perpendicular to the string. Note also that there is no
interaction term in the Lagrangian; instead for a pair of oscillators one
imposes the constraints:
\begin{equation}
x_{1}(t)=u(y_{1},t),\qquad x_{2}(t)=u(y_{2},t),  \label{ent2}
\end{equation}
where $y_{1}$ and $y_{2}$ are the positions of the oscillators along the
string. The equations of motion of the oscillators are
\begin{eqnarray}
m_{1}\ddot{x}_{1}+K_{1}x_{1} &=&f_{1}(t),  \notag \\
m_{2}\ddot{x}_{2}+K_{2}x_{2} &=&f_{2}(t),  \label{ent3}
\end{eqnarray}
where $f_{1}(t)$ and $f_{2}(t)$ are the constraint forces exerted by the
string on the individual oscillators. The field equation of motion for the
string becomes the inhomogeneous wave equation:
\begin{equation}
{\frac{\partial ^{2}u}{\partial t^{2}}}-c^{2}{\frac{\partial ^{2}u}{\partial
y^{2}}}=-{\frac{f_{1}(t)}{\sigma }}\delta (y-y_{1})-{\frac{f_{2}(t)}{\sigma }
}\delta (y-y_{2}),  \label{ent4}
\end{equation}
where $c=\sqrt{\tau /\sigma }$ is the wave velocity. The retarded solution
of this equation is
\begin{equation}
u(y,t)=u^{\text{h}}(y,t)-{\frac{1}{\zeta }}\int_{-\infty }^{t-\left\vert
y-y_{1}\right\vert /c}dt^{\prime }f_{1}(t^{\prime })-{\frac{1}{\zeta }}
\int_{-\infty }^{t-\left\vert y-y_{2}\right\vert /c}dt^{\prime
}f_{2}(t^{\prime }),  \label{ent5}
\end{equation}
where $u^{\text{h}}(y,t)$ is the operator solution of the homogeneous
equation corresponding to the free motion of the string in the absence of
the oscillators and
\begin{equation}
\zeta =2\sigma c=2\sqrt{\sigma \tau }  \label{ent6}
\end{equation}
is the friction constant. Differentiating with respect to $t$, we find
\begin{equation}
\frac{\partial u(y,t)}{\partial t}=\frac{\partial u^{\text{h}}(y,t)}{
\partial t}-{\frac{1}{\zeta }}f_{1}(t-\frac{\left\vert y-y_{1}\right\vert }{c
})-{\frac{1}{\zeta }}f_{2}(t-\frac{\left\vert y-y_{2}\right\vert }{c}).
\label{ent7}
\end{equation}
Putting first $y=y_{1}$ and then $y=y_{2}$, we find using the constraints (
\ref{ent2}) that we can write
\begin{eqnarray}
f_{1}\left( t\right) +f_{2}\left( t_{\text{R}}\right) &=&-\zeta \dot{x}
_{1}\left( t\right) +F_{1}\left( t\right) ,  \notag \\
f_{1}\left( t_{\text{R}}\right) +f_{2}\left( t\right) &=&-\zeta \dot{x}
_{2}\left( t\right) +F_{2}\left( t\right) ,  \label{ent8}
\end{eqnarray}
where $t_{\text{R}}$ is the retarded time,
\begin{equation}
t_{\text{R}}=t-\frac{\left\vert y_{1}-y_{2}\right\vert }{c},  \label{ent9}
\end{equation}
while $F_{1}(t)$ and $F_{2}(t)$ are fluctuating forces:
\begin{equation}
F_{1}(t)=\zeta \frac{\partial u^{\text{h}}(y_{1},t)}{\partial t},\quad
F_{2}(t)=\zeta \frac{\partial u^{\text{h}}(y_{2},t)}{\partial t}.
\label{ent10}
\end{equation}
We can solve the equations (\ref{ent8}) to find explicit expressions for $
f_{1}(t)$ and $f_{2}(t)$, but it is simpler to form the Fourier transforms
to get (remember $t_{\text{R}}$ is given by (\ref{ent9})
\begin{equation}
\left( 
\begin{array}{cc}
1 & e^{i\omega \left\vert y_{1}-y_{2}\right\vert /c} \\ 
e^{i\omega \left\vert y_{1}-y_{2}\right\vert /c} & 1
\end{array}
\right) \left( 
\begin{array}{c}
\tilde{f}_{1}(\omega ) \\ 
\tilde{f}_{2}(\omega )
\end{array}
\right) =i\omega \zeta \left( 
\begin{array}{c}
\tilde{x}_{1}(\omega ) \\ 
\tilde{x}_{2}(\omega )
\end{array}
\right) +\left( 
\begin{array}{c}
\tilde{F}_{1}(\omega ) \\ 
\tilde{F}_{2}(\omega )
\end{array}
\right) .  \label{ent11}
\end{equation}
If we form the Fourier transform of the equations of motion (\ref{ent3}) we
get
\begin{equation}
\left( 
\begin{array}{cc}
-m_{1}\omega ^{2}+K_{1} & 0 \\ 
0 & -m_{2}\omega ^{2}+K_{2}
\end{array}
\right) \left( 
\begin{array}{c}
\tilde{x}_{1}(\omega ) \\ 
\tilde{x}_{2}(\omega )
\end{array}
\right) =\left( 
\begin{array}{c}
\tilde{f}_{1}(\omega ) \\ 
\tilde{f}_{2}(\omega )
\end{array}
\right) .  \label{ent12}
\end{equation}
Eliminating the Fourier transform of the constraint forces between these two
equations, we can write
\begin{equation}
\left( 
\begin{array}{cc}
-m_{1}\omega ^{2}-i\omega \zeta +K_{1} & (-m_{2}\omega ^{2}+K_{2})e^{i\omega
\left\vert y_{1}-y_{2}\right\vert /c} \\ 
(-m_{1}\omega ^{2}+K_{1})e^{i\omega \left\vert y_{1}-y_{2}\right\vert /c} & 
-m_{2}\omega ^{2}-i\omega \zeta +K_{2}
\end{array}
\right) \left( 
\begin{array}{c}
\tilde{x}_{1}(\omega ) \\ 
\tilde{x}_{2}(\omega )
\end{array}
\right) =\left( 
\begin{array}{c}
\tilde{F}_{1}(\omega ) \\ 
\tilde{F}_{2}(\omega )
\end{array}
\right) .  \label{ent13}
\end{equation}
The susceptibility matrix is therefore
\begin{equation}
\mathbf{\alpha }(\omega )=\left( 
\begin{array}{cc}
-m_{1}\omega ^{2}-i\omega \zeta +K_{1} & (-m_{2}\omega ^{2}+K_{2})e^{i\omega
\left\vert y_{1}-y_{2}\right\vert /c} \\ 
(-m_{1}\omega ^{2}+K_{1})e^{i\omega \left\vert y_{1}-y_{2}\right\vert /c} & 
-m_{2}\omega ^{2}-i\omega \zeta +K_{2}
\end{array}
\right) ^{-1}.  \label{ent14}
\end{equation}
The quantum Langevin equations are
\begin{eqnarray}
m_{1}\ddot{x}_{1}(t)+\zeta \dot{x}_{1}(t)+K_{1}x_{1}(t)+m_{2}\ddot{x}_{2}(t_{
\text{R}})+K_{2}x_{2}(t_{\text{R}}) &=&F_{1}(t),  \notag \\
m_{2}\ddot{x}_{2}(t)+\zeta \dot{x}_{2}(t)+K_{2}x_{2}(t)+m_{1}\ddot{x}_{1}(t_{
\text{R}})+K_{1}x_{1}(t_{\text{R}}) &=&F_{2}(t).  \label{ent15}
\end{eqnarray}
The form of these equations makes clear that, at time $t$, there is an
interaction between $x_{1}=u(y_{1})$ and $x_{2}=u(y_{2})$. In particular, we
note that when there is an effect at $y_{1}$ due to the heat bath that it
does not effect $y_{2}$ until a later time $|y_{1}-y_{2}|/c$.

To get explicit expressions for the force correlations, we recall the well
known expression for the operator displacement of a string of length $L$:\cite{sakurai67} 
\begin{equation}
u^{\text{h}}\left( y,t\right) =\sum_{k}\sqrt{\frac{\hbar }{2\sigma L\omega }}
\left( a_{k}e^{i\left( ky-\omega t\right) }+a_{k}^{\dag }e^{-i\left(
ky-\omega t\right) }\right) ,  \label{ent16}
\end{equation}
where $\omega =c\left\vert k\right\vert $ and $k=\frac{2\pi n}{_{L}}$ with $
n=0,\pm 1,\pm 2,\cdots $. In equilibrium at temperature $T$, 
\begin{equation}
\left\langle a_{k}a_{k^{\prime }}^{\dag }+a_{k}^{\dag }a_{k^{\prime
}}\right\rangle =\coth \frac{\hbar \omega }{2k_{B}T}\delta _{k,k^{\prime }}.
\label{ent17}
\end{equation}
With this in the expressions (\ref{ent10}) for the fluctuating forces, we
find (in the limit $L\rightarrow \infty $) the correlations:
\begin{gather}
\frac{1}{2}\left\langle F_{1}(t)F_{1}(t^{\prime })+F_{1}(t^{\prime
})F_{1}(t)\right\rangle =\frac{1}{2}\left\langle F_{2}(t)F_{2}(t^{\prime
})+F_{2}(t^{\prime })F_{2}(t)\right\rangle  \notag \\
=\frac{\hbar \zeta }{\pi }\int_{0}^{\infty }d\omega \omega \coth \frac{\hbar
\omega }{2k_{B}T}\cos \left[ \omega \left( t-t^{\prime }\right) \right] 
\notag \\
=\frac{\zeta kT}{2}\frac{d}{dt}\coth \frac{\pi kT\left( t-t^{\prime }\right) 
}{\hbar },  \notag \\
\frac{1}{2}\left\langle F_{1}(t)F_{2}(t^{\prime })+F_{2}(t^{\prime
})F_{1}(t)\right\rangle =\frac{\hbar \zeta }{\pi }\int_{0}^{\infty }d\omega
\omega \coth \frac{\hbar \omega }{2k_{B}T}\cos \left[ \omega \left(
t-t^{\prime }\right) \right] \cos \frac{\omega \left( y_{1}-y_{2}\right) }{c}
\notag \\
=\frac{\zeta kT}{4}\frac{d}{dt}\left( \coth \frac{\pi kT\left( t-t^{\prime }-
\frac{y_{1}-y_{2}}{c}\right) }{\hbar }+\coth \frac{\pi kT\left( t-t^{\prime
}+\frac{y_{1}-y_{2}}{c}\right) }{\hbar }\right) .  \label{ent18}
\end{gather}
Here we should note in the explicit expressions that $d\coth \left( x\right)
/dx=2\delta (x)-$csch$^{2}(x)$ \cite{ford96,ford02}.

With these expressions for the correlations, we see from the Langevin
equations (\ref{ent15}) that, for large separations and times short compared
with $\left\vert y_{1}-y_{2}\right\vert /c$, the retarded time (\ref{ent6})
is in the distant past and the equations become decoupled. Each equation
becomes a Langevin equation for a single oscillator interacting with the
bath. There can still be correlation between the oscillators, for example
they can be in an entangled state, but there are no bath-induced
correlations.

The case of very small separations is more subtle. To simplify the
discussion we assume the oscillators are identical, that is, $m_{1}=m_{2}=m$
and $K_{1}=K_{2}=K$. Then by adding and subtracting the two Langevin
equations (\ref{ent15}) and introducing the variables:
\begin{equation}
X=\frac{x_{1}+x_{2}}{2},\quad x=x_{1}-x_{2},  \label{ent19}
\end{equation}
we obtain the separate equations \cite{ford85}
\begin{subequations}
\begin{eqnarray}
m\ddot{X}\left( t\right) +\zeta \dot{X}\left( t\right) +KX\left( t\right) +m
\ddot{X}\left( t_{\text{R}}\right) +KX\left( t_{\text{R}}\right) &=&\frac{
F_{1}\left( t\right) +F_{2}\left( t\right) }{2}, \\
m\ddot{x}\left( t\right) +\zeta \dot{x}\left( t\right) +Kx\left( t\right) -m
\ddot{x}\left( t_{\text{R}}\right) -Kx\left( t_{\text{R}}\right)
&=&F_{1}\left( t\right) -F_{2}\left( t\right) .  \label{ent20}
\end{eqnarray}
If we set $\left\vert y_{1}-y_{2}\right\vert =0$, then $t_{\text{R}}=t$.
Moreover, for very small separations, it is important to note that that all three of the correlations (
\ref{ent18}) of the fluctuation forces are the same; there is only one
fluctuation force. Therefore the first of the equations (\ref{ent20})
becomes exactly the Langevin equation for a single oscillator whose mass is
the sum of the masses and with a spring constant equal to the sum of the
spring constants. This is what we should expect, when the two oscillators
merge they should become one. But what about the second of the equations (
\ref{ent20})? With $\left\vert y_{1}-y_{2}\right\vert =0$ this equation
becomes 
\end{subequations}
\begin{equation}
\zeta \dot{x}\left( t\right) =0.  \label{ent21}
\end{equation}
with no fluctuating force. Thus $x=x_{1}-x_{2}$ can take on any constant
value!

To better understand this strange result we consider the free energy for the
system, which is given by the remarkable formula:\cite{ford85,li90}
\begin{equation}
F\left( \left\vert y_{1}-y_{2}\right\vert ,T\right) =\frac{1}{\pi }
\int_{0}^{\infty }d\omega f\left( \omega ,T\right) \text{\textrm{Im}}\left\{ 
\frac{d\log \det \mathbf{\alpha }\left( \omega +i0^{+}\right) }{d\omega }
\right\} ,  \label{ent22}
\end{equation}
where $\mathbf{\alpha }$ is the polarizability matrix (\ref{ent14}) and
\begin{equation}
f\left( \omega ,T\right) =kT\log \left( 2\sinh \frac{\hbar \omega }{2kT}
\right)  \label{ent23}
\end{equation}
is the free energy of a single oscillator of natural frequency $\omega $. I
now we put $\left\vert y_{1}-y_{2}\right\vert =0$ in the polarizability
matrix (\ref{ent14}) we find
\begin{equation}
\det \mathbf{\alpha }\left( \omega \right) =\frac{1}{-i\omega \zeta }\times 
\frac{1}{-\left( m_{1}+m_{2}\right) \omega ^{2}+K_{1}+K_{2}-i\omega \zeta }.
\label{ent24}
\end{equation}
Here the second factor is exactly the polarizability of a single oscillator
with mass the sum of the masses and spring constant equal to the sum of the
spring constants. When put in the remarkable formula this factor will give
the corresponding free energy. \ The first factor will give rise to an
additional free energy
\begin{equation}
\Delta F\left( 0,T\right) =\frac{1}{2}f\left( 0,T\right) ,  \label{ent25}
\end{equation}
where we have used the well known identity: \textrm{Im}$\left\{ 1/\left(
\omega +i0^{+}\right) \right\} =-\pi \delta \left( \omega \right) $ and the
factor of $\frac{1}{2}$ takes into account the fact that the delta function
is at the initial point of the integration. This additional free energy is
infinite: an oscillator whose natural frequency is zero is an unbounded free
particle. We conclude that, in the limit $\left\vert y_{1}-y_{2}\right\vert
\rightarrow 0$, the free energy of the system is infinite, but that the
divergent part can be isolated and remainder is finite and can be recognized
as the free energy of a single oscillator coupled to the bath.

We now turn to a calculation of the fluctuation force between two masses
attached to a string. In order to calculate the stress between the two
masses, we must first obtain the solution of the equation of motion of the
string $u\left( y,t\right) $ corresponding to an external force $f\left(
y,t\right) $, in the form of an integral involving the Green function $
G\left( y,y^{\prime };t\right) $. We then solve for the Fourier transform of
the Green function at points between the two masses, which will involve
reflection coefficients $r_{1}$ at $y_{1}$ and $r_{2}$ at $y_{2}$. Next, the
stress at a point between the two masses is calculated in terms of space and
time derivatives of the solution of the equations of motion, which leads to
an expression in terms of the spatial correlation function, which in turn is
evaluated by means of the fluctuation-dissipation theorem in terms of the
Green function.

Thus, we start with two equal masses $m$ attached to the string, one at $
y_{1}$ the other at $y_{2}$. The Green function is used to express the
solution of the equation of motion of the string corresponding to an
external force $f\left( y,t\right) $ in the form
\begin{equation}
u\left( y,t\right) =\int_{-\infty }^{t}dt^{\prime }\int_{-\infty }^{\infty
}dy^{\prime }G\left( y,y^{\prime }:t-t^{\prime }\right) f\left( y,^{\prime
},t^{\prime }\right) .  \label{ent26}
\end{equation}
Clearly the Green function $G\left( y,y^{\prime };t\right) $ that vanishes
for negative times corresponds to an external force $\delta \left(
y-y^{\prime }\right) \delta (t)$. For a source point $y^{\prime }$ and field
point $y$ both in the interval between the two masses, the Fourier transform
of the Green function is a solution of the inhomogeneous Helmholtz equation:
\begin{equation}
\frac{\partial ^{2}\tilde{G}\left( y,y^{\prime };\omega \right) }{\partial
y^{2}}+\frac{\omega ^{2}}{c^{2}}\tilde{G}\left( y,y^{\prime };\omega \right)
=-\frac{1}{\tau }\delta \left( y-y^{\prime }\right) .  \label{ent27}
\end{equation}
Still in the region between the pair of masses $\left( y_{2}<y,y^{\prime
}<y_{1}\right) $ the Green function for $y>y^{\prime }$ has the form of a
plane wave reflecting off the mass at $y_{1}$, while for $y<y^{\prime }$ it
has the form of a plane wave reflecting off the mass at $y_{2}$. In addition
to satisfy the inhomogeneous equation, $\tilde{G}\left( y,y^{\prime };\omega
\right) $ must be continuous and have a jump of $-1/\tau $ in its derivative
at $y=y^{\prime }$. It follows that 
\begin{eqnarray}
&&\tilde{G}\left( y,y^{\prime };\omega \right) =  \notag \\
&=&\frac{i}{2k\tau }\left\{ 
\begin{array}{c}
\frac{\left[ e^{ik\left( y_{2}-y^{\prime }\right) }+r_{2}e^{-ik\left(
y_{2}-y^{\prime }\right) }\right] \left[ e^{ik\left( y-y_{1}\right)
}+r_{1}e^{-ik\left( y-y_{1}\right) }\right] }{e^{-ik\left(
y_{1}-y_{2}\right) }-r_{1}r_{2}e^{ik\left( y_{1}-y_{2}\right) }},\qquad
y^{\prime }<y<y_{1} \\ 
\frac{\left[ e^{ik\left( y^{\prime }-y_{1}\right) }+r_{1}e^{-ik\left(
y^{\prime }-y_{1}\right) }\right] \left[ e^{ik\left( y_{2}-y\right)
}+r_{2}e^{-ik\left( y_{2}-y\right) }\right] }{e^{-ik\left(
y_{1}-y_{2}\right) }-r_{1}r_{2}e^{ik\left( y_{1}-y_{2}\right) }},\qquad
y_{2}<y<y^{\prime }
\end{array}
\right. ,  \label{ent28}
\end{eqnarray}
where $r_{1}$ and $r_{2}$ are the reflection coefficients at $y_{1}$ and $
y_{2}$, respectively. Note that the Green function is symmetric under
interchange of $y$ and $y^{\prime }$: $\tilde{G}\left( y^{\prime },y;\omega
\right) =\tilde{G}\left( y,y^{\prime };\omega \right) $.

Next, we note that the mean of the stress $T$ at a point between the masses
is equal to the mean of the energy density:
\begin{eqnarray}
\left\langle T\right\rangle &=&\left\langle \frac{1}{2}\sigma (\frac{
\partial u}{\partial t})^{2}+\frac{1}{2}\tau (\frac{\partial u}{\partial y}
)^{2}\right\rangle  \notag \\
&=&\left[ \left( \frac{1}{2}\sigma \frac{\partial ^{2}}{\partial t\partial
t^{\prime }}+\frac{1}{2}\tau \frac{\partial ^{2}}{\partial y\partial
y^{\prime }}\right) C\left( y,y^{\prime };t-t^{\prime }\right) \right] 
_{\substack{ t^{\prime }=t  \\ y^{\prime }=y}},  \label{ent29}
\end{eqnarray}
where 
\begin{equation}
C\left( y,y^{\prime };t-t^{\prime }\right) =\frac{1}{2}\left\langle u\left(
y,t\right) u\left( y^{\prime },t^{\prime }\right) +u\left( y^{\prime
},t^{\prime }\right) y\left( x,t\right) \right\rangle  \label{ent30}
\end{equation}
is the correlation function of the string displacement (only $t-t^{\prime }$
appears because of time translation invariance). The fluctuation-dissipation
theorem relates this correlation function to the Green function: 
\begin{equation}
\tilde{C}\left( y,y^{\prime };\omega \right) =\frac{\hbar }{2i}\coth \frac{
\hbar \omega }{2k_{B}T}\left[ \tilde{G}\left( y,y^{\prime };\omega \right) -
\tilde{G}\left( y^{\prime },y;\omega \right) ^{\ast }\right] .  \label{ent31}
\end{equation}
Therefore 
\begin{equation}
\left\langle T\right\rangle =\frac{\hbar }{2\pi }\int_{-\infty }^{\infty
}d\omega \coth \frac{\hbar \omega }{2k_{B}T}\left[ \left( \frac{1}{2}\sigma
\omega ^{2}+\frac{1}{2}\tau \frac{\partial ^{2}}{\partial y\partial
y^{\prime }}\right) \text{\textrm{Im}}\left\{ \tilde{G}\left( y,y^{\prime
};\omega \right) \right\} \right] _{y^{\prime }=y}.  \label{ent32}
\end{equation}
Thus, using equation (\ref{ent28}) for the Green function we get 
\begin{equation}
\left\langle T\right\rangle =\frac{\hbar }{4\pi c}\int_{-\infty }^{\infty
}d\omega \omega \coth \frac{\hbar \omega }{2k_{B}T}\text{\textrm{Re}}\left\{ 
\frac{1+r_{1}r_{2}e^{2ik|y_{1}-y_{2}|}}{1-r_{1}r_{2}e^{2ik|y_{1}-y_{2}|}}
\right\} .  \label{ent33}
\end{equation}
We interpret the fluctuation force between the masses as the difference
between the stress in the presence of the attached masses and the free space
stress ($r_{1}=r_{2}=0$). That is,
\begin{eqnarray}
F &=&\left\langle T\right\rangle -\left\langle T\right\rangle _{0}  \notag \\
&=&\frac{\hbar }{2\pi c}\int_{-\infty }^{\infty }d\omega \omega \coth \frac{
\hbar \omega }{2k_{B}T}\mathrm{Re}\left\{ \frac{r_{1}r_{2}e^{2ik\left\vert
y_{1}-y_{2}\right\vert }}{1-r_{1}r_{2}e^{2ik\left\vert
y_{1}-y_{2}\right\vert }}\right\} .  \label{ent34}
\end{eqnarray}

To proceed we must determine the reflection coefficients. To do so we
consider scattering by a single mass at $y=0$, for which the scattering wave
is of the form.
\begin{equation}
u\left( y,t\right) =e^{-i\omega t}\left\{ 
\begin{array}{c}
e^{iky}+re^{-iky},\qquad y<0 \\ 
te^{iky},\qquad y>0
\end{array}
\right.  \label{ent35}
\end{equation}
where $r$ and $t$ are the reflection and transmission coefficients, while $
k=\omega /c$. At $y=0$ we must require continuity ($\left[ u\right] =0$) and
a jump in the derivative corresponding to the force exerted by the mass ($
\left[ \tau \partial u/\partial y\right] =-m\omega ^{2}u$) The result is a
pair of equations for the determination of $r$ and $t$. Their solution is
\begin{equation}
t=\frac{\zeta }{\zeta -im\omega },\qquad r=\frac{im\omega }{\zeta -im\omega }
,  \label{ent36}
\end{equation}
where $\zeta $ is given in Eq. (\ref{ent18}). Therefore, since the masses
are assumed identical, in the expression (\ref{ent34}) for the fluctuation
force we set $r_{1}=r_{2}=r$, given by the above expression, to get
\begin{equation}
F=\frac{\hbar }{2\pi c}\mathrm{Re}\left\{ \int_{-\infty }^{\infty }d\omega
\omega \coth \frac{\hbar \omega }{2k_{B}T}\frac{\left( \frac{im\omega }{
\zeta -im\omega }\right) ^{2}e^{2ik\left\vert y_{1}-y_{2}\right\vert }}{
1-\left( \frac{im\omega }{\zeta -im\omega }\right) ^{2}e^{2ik\left\vert
y_{1}-y_{2}\right\vert }}\right\} .  \label{ent37}
\end{equation}
In the integral appearing here, the only singularities of the integrand in
the upper half plane are poles of the hyperbolic cotangent at $\omega
=in\Omega ,\quad n=1,2,\cdots $, where
\begin{equation}
\Omega =\frac{2\pi k_{B}T}{\hbar }  \label{ent38}
\end{equation}
is the Matsubara frequency.\ Therefore, we can deform the contour of
integration into the upper half plane, picking up the contribution of these
poles, and write
\begin{equation}
F=-\frac{\hbar \Omega ^{2}}{\pi c}\sum_{n=1}^{\infty }n\frac{\left( \frac{
nm\Omega }{\zeta +nm\Omega }\right) ^{2}e^{-2n\Omega \left\vert
y_{1}-y_{2}\right\vert /c}}{1-\left( \frac{nm\Omega }{\zeta +nm\Omega }
\right) ^{2}e^{-2n\Omega \left\vert y_{1}-y_{2}\right\vert /c}}.
\label{ent39}
\end{equation}
In the low temperature limit $\Omega \rightarrow 0$ and we can replace the
sum by an integral to write
\begin{equation}
F=-\frac{\hbar c}{\pi \left\vert y_{1}-y_{2}\right\vert ^{2}}
\int_{0}^{\infty }dww\frac{\left( we^{-w}\right) ^{2}}{\left( \frac{\zeta
\left\vert y_{1}-y_{2}\right\vert }{mc}+w\right) ^{2}-\left( we^{-w}\right)
^{2}}.  \label{ent40}
\end{equation}
This is the result for the fluctuation force at zero temperature. In the
limits of large and small separation the integrals are standard, and we
obtain:
\begin{equation}
F=-\left\{ 
\begin{array}{c}
\frac{\pi \hbar c}{24\left\vert y_{1}-y_{2}\right\vert ^{2}},\quad \frac{
\zeta \left\vert y_{1}-y_{2}\right\vert }{mc}\ll 1 \\ 
\frac{3m^{2}\hbar c^{3}}{8\pi \zeta ^{2}\left\vert y_{1}-y_{2}\right\vert
^{4}}\quad \frac{\zeta \left\vert y_{1}-y_{2}\right\vert }{mc}\gg 1
\end{array}
\right.  \label{ent41}
\end{equation}
The result for small separation is the analog for our system of the London
formula\cite{london26} for the force between neutral atoms, the large
separation result is the analog of the Casimir-Polder\cite{casimir48}
formula.

Our aim has been to show that there is a rich variety of phenomena and
plenty of unanswered questions associated with the problem of a pair of
oscillators coupled to a common heat bath. But more we have tried to show
that one must have a model of the bath for which there is a clear notion of
the separation of the oscillators. For this reason we chose to discuss the
simple Lamb model of a stretched string. Another obvious model would be the
blackbody bath. Indeed, the Lamb model is sometimes called scalar
electrodynamics and our calculation of the force between a pair of masses
attached to the string is a toy example of the Lifshitz calculation of the
Casimir force between two plates. Finally, we have strived to caution
against a discussion of the problem based on the independent oscillator
model. We have been exponents of that model because of its great generality,
\cite{ford88} but in its general form there is no notion of separation. The
models we have advocated, the Lamb model and the blackbody bath, are special
cases of independent oscillator models formulated with a clear notion of the
distance between the oscillators.

The conclusion we would draw is that the problem of two or more oscillators
in a common heat bath can be formulated and some of the interesting
questions answered, but not from an independent oscillator model.

This work was supported in part by the National Science Foundation under
grant no. ECCS-1125675.

\end{document}